\documentstyle[epsfig,twocolumn]{venice97}

\begin{document}

%%%%%%%%%%%%%%%%%%%%%%%%%%%%%%%%%%%%%%%%%%%%%%%%%%%%%%%%%%
%
% USER-DEFINED MACROS
%
\def\spose#1{\hbox to 0pt{#1\hss}}
\def\la{\mathrel{\spose{\lower 3pt\hbox{$\mathchar"218$}}
     \raise 2.0pt\hbox{$\mathchar"13C$}}}
\def\ga{\mathrel{\spose{\lower 3pt\hbox{$\mathchar"218$}}
     \raise 2.0pt\hbox{$\mathchar"13E$}}}
%
%%%%%%%%%%%%%%%%%%%%%%%%%%%%%%%%%%%%%%%%%%%%%%%%%%%%%%%%%%

\setlength{\parindent}{0pt}
\setlength{\parskip}{10pt plus 1pt minus 1pt}
\setlength{\hoffset}{-1.5truecm}
\setlength{\textwidth}{17.1truecm}
\setlength{\columnsep}{1truecm }
\setlength{\columnseprule}{0pt}
\setlength{\headheight}{12pt}
\setlength{\headsep}{20pt}
\pagestyle{veniceheadings}

\title{\bf IMPROVED METHODS FOR IDENTIFYING MOVING GROUPS}

\author{{\bf
	J.H.J.~de Bruijne$^1$,
	R.~Hoogerwerf$^1$,
	A.G.A.~Brown$^1$,
	L.A.~Aguilar$^2$,
	P.T.~de Zeeuw$^1$}
\vspace{2mm} \\
$^1$ Sterrewacht Leiden, P.O.\ Box 9513, 2300 RA Leiden, The Netherlands \\
$^2$ Instituto de Astronom\'{\i}a, U.N.A.M., P.O.\ Box 877, Ensenada,
     22830 Baja California, Mexico}

\maketitle

\begin{abstract}

We present a new procedure for the identification of moving groups. It
is a combination of two independent member selection methods. One is a
modern implementation of a classical convergent point method for
proper motion data. The other objectively identifies moving groups in
velocity space using proper motions {\em and} parallaxes. We briefly
describe both methods, and illustrate their power by applying them to
Hipparcos measurements of a field containing the Upper Scorpius
subgroup of the Scorpio--Centaurus association (Sco OB2). We show how
our membership selection procedure not only improves the list of
previously known B and A-type members, but also identifies many new
members, including a significant number of F stars.  We apply our
procedure to other nearby OB associations elsewhere in these
proceedings (Hoogerwerf et al.; de Zeeuw et al.).
\vspace {5pt} \\


\end{abstract}

\section{INTRODUCTION}
\label{introduction}

Stars in moving groups with small internal velocity dispersions share
a common space motion, which can be used to establish membership based
on measurements of radial velocities or proper motions, or both.
Whereas many such kinematic membership studies have been carried out
for open clusters (e.g., van Leeuwen 1985; van Altena et al.\ 1993),
there are few such studies for nearby OB associations.  These loose
stellar groups have internal velocity dispersions of at most a few
km/sec (cf.\ Mathieu 1986), so the stars share a common space motion,
but they generally cover tens to hundreds of square degrees on the
sky. Ground-based proper motion studies therefore almost invariably
have been confined to modest samples of bright stars ($V \la 6$) in
fundamental catalogs, or to small areas covered by a single
photographic plate. Photometric studies can extend membership to later
spectral types, but are less reliable. As a result, membership for
many associations has previously been determined unambiguously only
for spectral types earlier than B5 (e.g., Blaauw 1991).

The advent of Hipparcos allows a major step forward in our
understanding of the nearby associations. We are carrying out a census
of OB associations within 800~pc from the Sun based on the Hipparcos
measurements (see Hoogerwerf et al.\ 1997; de Zeeuw et al.\ 1997). The
parallaxes and proper motions allow kinematic membership determination
in the nearby associations to unprecedented accuracy, and for stars
significantly fainter than accessible previously.  This requires an
objective and reliable way to identify moving groups in the Hipparcos
Catalog. Here we present a new procedure to do just this, and
illustrate its application by considering the well-studied Upper
Scorpius subgroup of the Scorpio--Centaurus association (e.g., Blaauw
1978).

\section{SELECTION METHODS}
\label{selection_methods}

\begin{figure}[!b]
%%%%%%%%%%%%%%%%%%%%
%
% Spaghetti figure
%
%%%%%%%%%%%%%%%%%%%%
\begin{center}
\leavevmode
\centerline{\epsfig{file=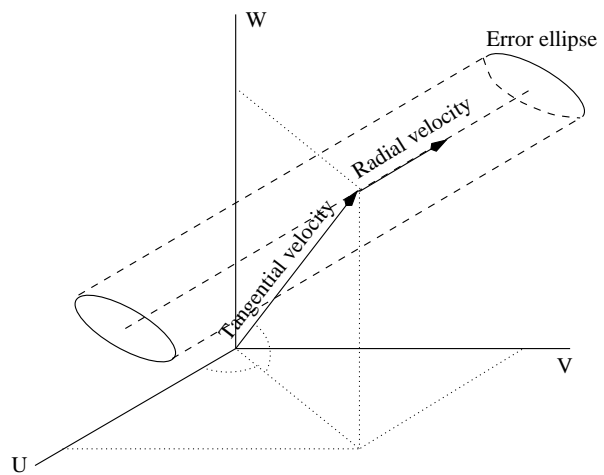,angle=270,width=8.0cm}}
\end{center}
\caption{\em
The five astrometric parameters of a star measured by Hipparcos define
an elliptic cylinder in velocity space. All cylinders of stars in a
moving group intersect.
}
\label{fig1}
\end{figure}

\begin{figure*}[!ht]
%%%%%%%%%%%%%%%%%%%%
%
% Grand data figure
%
%%%%%%%%%%%%%%%%%%%%
\begin{center}
\leavevmode
\centerline{\epsfig{file=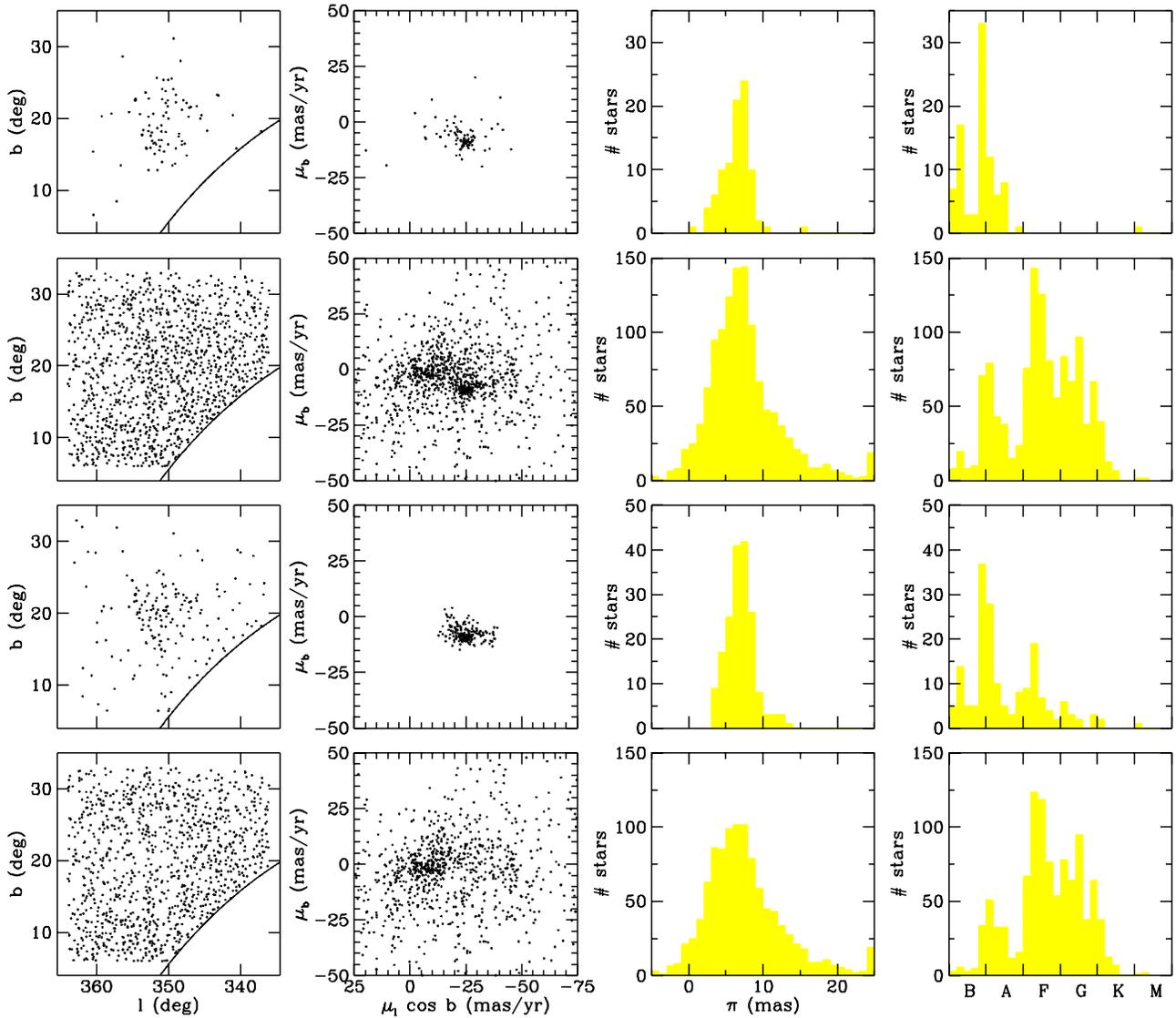,width=\textwidth}}
\end{center}
\caption{\em
Hipparcos measurements for Upper Scorpius (from the top row down): (1)
the 91 pre-Hipparcos members; (2) our data sample; (3) the 178
Hipparcos members; (4) the remaining stars in the sample after
membership selection.  The columns show (from left to right): (1)
positions in Galactic coordinates; (2) Galactic vector point diagram;
(3) trigonometric parallax distribution; (4) spectral type
distribution.  The M-type member is Antares ($\alpha$ Sco, M1Ib). The
solid lines in column (1) denote the boundary $\delta = -34^{\circ}$
with Upper Centaurus Lupus.  In the second panel of column (2), the
Upper Scorpius clump around $(\mu_{l} \cos b, \mu_{b}) \approx (-25,
-10)$~mas/yr is seen superimposed on the Galactic disk
distribution. The spread in this clump is caused by a combination of
perspective effects and observational errors.
}
\label{fig2}
\end{figure*}

The common space motion of stars in a nearby moving group results in
converging proper motions.  We developed a modern implementation of a
classical convergent point method (Brown 1950; Jones 1971). This
method searches for the maximum likelihood coordinates $(\alpha,
\delta)_{\rm cp}$ of the convergent point by minimizing the sum over
all $N$ stars in the data sample of the squared and error-weighted
proper motion components $\mu_{\perp} / \sigma_{\perp}$ perpendicular
to the direction to the corresponding convergent point.  This sum is
distributed as $\chi^2$ with $N \! - \! 2$ degrees of freedom.  If,
after global minimization with respect to $(\alpha, \delta)_{\rm cp}$,
the value of $\chi^2$ is unacceptably high, the star with the highest
value of $\mu_{\perp} / \sigma_{\perp}$ is rejected, after which
minimization is repeated until a satisfactory value of $\chi^2$ is
obtained.  Subsequently, all non-rejected stars are identified as
members.  This procedure allows for simultaneous convergent point
determination and membership selection.

\begin{figure*}[!th]
%%%%%%%%%%%%%%%%%%%%
%
% Members as function of spectral type figure
%
%%%%%%%%%%%%%%%%%%%%
\begin{center}
\leavevmode
\centerline{\epsfig{file=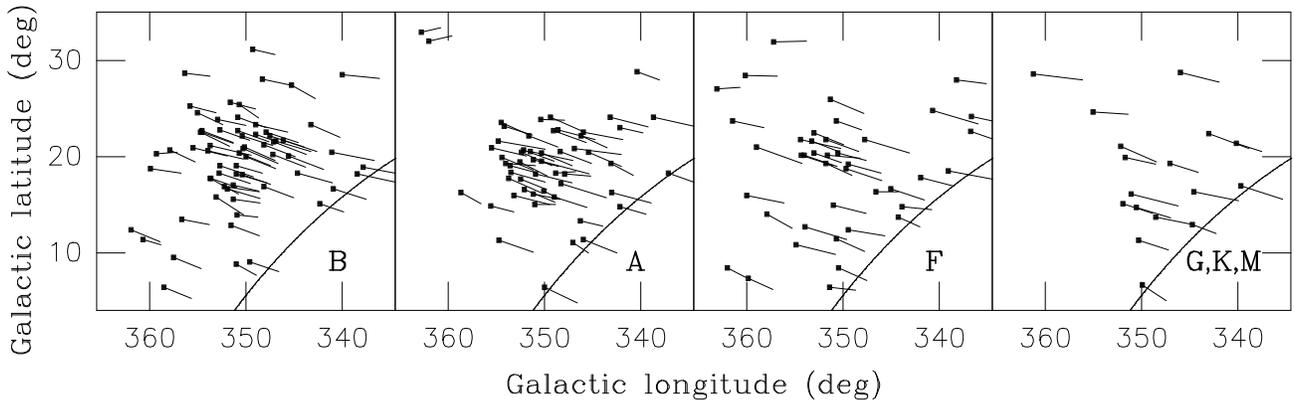,width=\textwidth}}
\end{center}
\caption{\em
Proper motions for the Hipparcos members of Upper Scorpius, as a
function of spectral type: 66 B, 54 A, 41 F, 14 G, 2 K stars, and 1 M
star.  The F, G, K, and M stars are co-moving with the B and A
stars. The F-type stars have a similar distribution as the early-type
stars, suggesting that the stellar content of the association has been
established down to at least spectral type F.  The probability that
many of the G-type stars are interloper field stars is significant.
Some stars near the field boundary $\delta = -34^\circ$ may be members
of Upper Centaurus Lupus. 
}
\label{fig3}
\end{figure*}

In order to make optimal use of the Hipparcos data, we developed a new
kinematic membership selection method which uses, besides proper
motions, also parallaxes. In this method, each star is characterized
by a line in velocity space: the proper motion and parallax determine
the offset from the origin (tangential velocity), while the sky
position of the star determines its direction (radial velocity; cf.\
Figure~\ref{fig1}). The errors and correlations of the astrometric
parameters transform the line into an elliptic cylinder.  The
cylinders of a set of stars with the same space motion all intersect.
Thus, we can identify moving groups by searching for maxima in the
density of cylinders in velocity space. All stars whose 3 sigma
cylinders contain a maximum are selected as member of the associated
group. The membership probability depends on the velocity dispersion
of the group, and on the errors in the measurements.

We have tested both methods on the Hyades cluster.  Based on Hipparcos
data and radial velocities, Perryman et al.\ (1997) identified 218
members which have three-dimensional space motions consistent with the
cluster motion within 3$\sigma$ limits.  Within 10~pc of the cluster
center ($\sim r_{\rm t}$, the tidal radius), our membership list
agrees very well with that of Perryman et al.  Furthermore, we have
performed extensive tests of both methods based on numerically
simulated data sets of moving groups superimposed on a kinematic model
of the Galactic disk.  We have varied the position of the group on the
sky, as well as its distance, streaming motion, internal velocity
dispersion, and number of members. The observational errors and
covariance matrix for each simulated star were chosen to be consistent
with the data in the Hipparcos Catalog.  These tests indicate, in
order of decreasing importance, that (i) a clear identification of the
moving group requires it to stand out kinematically from the Galactic
field distribution; and (ii) an increasing distance of the moving
group complicates its identification.  When proper motions combined
with parallaxes do not allow the separation of the moving group from
the Galactic field, accurate radial velocities are an indispensable
tool in establishing kinematic membership (cf.\ the case of Orion OB1
discussed by de Zeeuw et al.\ 1997)

Both methods take the full Hipparcos covariance matrix into account.
Neither method requires a priori knowledge on the (existence or
characteristics of the) moving group. We combine the independent
results of the two methods to define membership criteria: we consider
as secure members all stars that are selected by both methods. Here we
restrict our discussion to these stars.

\section{UPPER SCORPIUS}
\label{application}

Classical kinematic membership studies for Upper Scorpius identified
59 proper motion members with spectral types earlier than B8 (Blaauw
1946; Bertiau 1958; Jones 1971).  Several photometric studies
suggested membership for another 32 stars down to spectral type A8
(Hardie \& Crawford 1961; Garrison 1967; Gutierrez--Moreno \& Moreno
1968; Glaspey 1971, 1972; de Geus et al.\ 1989). The upper panels of
Figure~\ref{fig2} display the Hipparcos measurements for these 91
pre-Hipparcos members.

Our Hipparcos data sample for Upper Scorpius (cf.\ second panel row of
Figure~\ref{fig2}) is limited to the field
$336^{\circ} \leq \ell \leq  4^{\circ}$,
$  6^{\circ} \leq    b \leq 33^{\circ}$,
with the extra restriction $\delta \geq -34^{\circ}$, which provides a
boundary with the subgroup Upper Centaurus Lupus of Sco OB2 (cf.\ de
Zeeuw et al.\ 1997). The sample contains 1215 stars within broad
magnitude ranges, which depend on spectral type (cf.\ 
Table~\ref{table1}).

\begin{table}[!b]
\caption{\em
Visual magnitude limits of the Hipparcos data sample for Upper
Scorpius, as function of spectral type.
}
\label{table1}
\begin{center}
\leavevmode
\footnotesize
\begin{tabular}[h]{rrrrrr}
\hline \\[-5pt]
Spec.\ type & Range & \# & Spec.\ type & Range & \# \\[+5pt]
\hline \\[-5pt]
O  -- B5 &       -- $7.0$ &  31 & F0 -- F4 & $7.5$ -- & 238  \\
B6 -- B9 &       -- $8.0$ &  86 & F5 -- F9 & $8.0$ -- & 244  \\
A0 -- A4 & $5.0$ -- $9.0$ & 130 & G        & $8.5$ -- & 353  \\
A5 -- A9 & $7.0$ -- $9.5$ &  69 & K, M     &          &  64  \\[+5pt]
\hline
\end{tabular}
\end{center}
\end{table}

We first apply our membership selection to the B-type stars. This
results in 66 members. The same procedure applied to the A-type stars
yields 54 members.  Application of our procedure to the {\em combined}
sample of B and A-type stars gives identical results.  We also detect
the association in the F-type stars, but not as convincing as in the
early-type stars.  We find no signature of a moving group in the
G-type stars.  Therefore, we decided to use the 120 B and A-type
members to define the space motion of Upper Scorpius.  Then, we select
the F, G, K, and M-type stars that have a space motion consistent with
that of the early-type members.  This results in 58 co-moving members.
The third panel row of Figure~\ref{fig2} shows the measurements for
all 178 Hipparcos members; 63 of them were previously known (43
kinematic, 20 photometric), while 115 of them are new! We confirm
Antares ($\alpha$ Sco, M1Ib) as member, but reject the controversial
classical proper motion member o Sco (A4II/III; see Blaauw et al.\
1955; de Geus et al.\ 1989; de Zeeuw et al.\ 1997: figure 2).  The
bottom panel row of Figure~\ref{fig2} shows the remaining field star
population after membership selection.  The small hole in the vector
point diagram around $(\mu_{l} \cos b, \mu_{b}) \approx (-20,
-10)$~mas/yr indicates that a few field stars might erroneously have
ended up as members (see discussion below).

Analysis of the parallax distribution results in a mean distance $d =
145 \pm 2$~pc (where the quoted error corresponds to the error in the
mean parallax), which can be compared to the value of $d = 160 \pm
40$~pc obtained by de Geus et al.\ (1989). The intrinsic depth of the
association is not resolved by the Hipparcos parallaxes. The observed
large spread in the vector point diagram is consistent with zero
internal velocity dispersion. It can be fully explained by the
combined effects of observational errors and projection on the
sky. Unfortunately, this implies that the study of internal motions in
associations based on Hipparcos proper motions is beyond our current
capabilities.

Figure~\ref{fig3} shows the positions and proper motions for all the
Hipparcos members, as a function of spectral type. The figure shows
that there is a clear concentration in the early-type stars,
surrounded by a group of `outliers'.  However, all stars in
Figure~\ref{fig3} have astrometric parameters that are consistent with
the space motion of the subgroup. We can remove interlopers (field
stars) by using the radial velocity to determine the three-dimensional
space motion for each star, and comparing it to that of the
subgroup. In the absence of a coherent and homogeneous set of radial
velocities for the majority of stars in our list of Hipparcos members,
we have estimated the number of expected interlopers by Monte--Carlo
simulation. We find that the observed clustering in the F-type star
distribution cannot be explained by Galactic disk contamination,
whereas the majority of the G-type stars can be explained as such.

The main concentration of F-type stars has spectral types earlier than
$\sim$F5. The Hayashi time scale for mid-F type stars is of the order
of 5~Myr, which is comparable to the age of Upper Scorpius estimated
from the Hertzsprung--Russell diagram (e.g., de Geus et al.\ 1989).
If stars of later spectral types have formed in Upper Scorpius they
should lie above the main sequence. As these objects lie near the
faint limit of the Hipparcos Catalog, careful analysis is needed to
establish their numbers. 

\section{CONCLUDING REMARKS}
\label{conclusions}

We have developed a new method for identifying moving groups in
velocity space, based on all five astrometric parameters obtained by
Hipparcos.  Combination of this method with a modern version of a
classical convergent point method leads to a powerful procedure for
identifying moving groups in the Hipparcos Catalog. We applied this
procedure to Hipparcos data for 21 fields related to the nearby OB
associations. Results for some of these fields are described by
Hoogerwerf et al.\ (1997) and de Zeeuw et al.\ (1997), the latter of
which also gives a general overview of the project.  Here we present
results for Upper Scorpius. We establish a significant improvement of
the pre-Hipparcos membership list, and an extension of it to much
later spectral types including F-type stars, for a total of 178
kinematic members.  Our membership list represents a suitable working
set of stars; radial velocities are needed to remove remaining
interlopers. The Hipparcos proper motions do not allow an analysis of
the internal kinematics of Upper Scorpius.

\section*{ACKNOWLEDGMENTS}

It is a pleasure to thank Adriaan Blaauw, Eug\`{e}ne de Geus and
Michael Perryman for many stimulating discussions, and the latter for
providing membership data of the Hyades in advance of publication.

\end{document}